\documentclass[12pt]{article}

\usepackage{amsmath,amssymb}
\usepackage{epsfig,graphicx}

\newcommand{\p}{\bot}
\newcommand{\dd}{\partial}
\newcommand{\de}{\delta}
\newcommand{\om}{\omega}
\newcommand{\e}{\varepsilon}
\newcommand{\ls}{\left(}
\newcommand{\rs}{\right)}
\newcommand{\g}{\gamma}
\newcommand{\m}{\mu}
\newcommand{\n}{\nu}
\newcommand{\ra}{\rangle}
\newcommand{\te}{\theta}

\newcommand{\eel}{e}
\newcommand{\f}{\phi}

\newcommand{\disn}[2]{$$\displaylines{\refstepcounter{equation}%
            \label{#1}\hskip 1em minus 1em #2\hfilneg}$$}
\newcommand{\nom}{\hfil\hskip 1em minus 1em (\theequation)}
\newcommand{\no}{\hfil \hskip 1em minus 1em\phantom{(\theequation)}%
            \hfilneg\cr\hfilneg\hskip 1em minus 1em\hfil}
\newcommand{\ns}{\hfill\cr\hfill}

\makeatletter

\long\def\@makecaption#1#2{%
   \vskip 10\p@
   \setbox\@tempboxa\hbox{#1. #2}%
   \ifdim \wd\@tempboxa >\hsize
       #1. #2\par
     \else
       \hbox to\hsize{\hfil\box\@tempboxa\hfil}%
   \fi}

\renewcommand{\section}{\@startsection{section}{1}{0pt}%
          {3.5ex plus 1ex minus .2ex}{2.3ex plus .2ex}{\noindent\hfil\bf}}

\makeatother

\begin{document}
\title{
QED(1+1) on the Light Front
and its implications\\
for semiphenomenological methods in QCD(3+1)\\
}
\author{V.A.~Franke, S.A.~Paston, E.V.~Prokhvatilov\\
{\it St.-Petersburg State University, Russia}
}
\date{\vskip 15mm}
\maketitle

\begin{abstract}
A possibility of semiphenomenological description of vacuum effects
in QCD quantized on the  Light Front (LF)  is discussed. A modification of
the canonical LF Hamiltonian for QCD is proposed, basing on
the detailed study of the exact description of
vacuum condensate  in QED(1+1)  that uses correct form of LF Hamiltonian.
\end{abstract}

\newpage
\section{Introduction}
First of all let us remind briefly basic advantages and main difficulties
of the quantization on the LF (by the "LF" we mean the hyperplane $x^+=0$ in
Dirac \cite{dir}
"light cone" coordinates $x^{\pm}=(x^0\pm x^1)/\sqrt{2}$, with the
$x^+$  playing the role of time).

\vskip 3mm
{\bf 1.} LF momentum operator $P_-=(P_0-P_1)/\sqrt{2}\ge 0$
is nonnegative for states with $p_0\ge0$, $p^2\ge0$.
Like usual space momentum it is kinematical
(quadratic in fields) generator of translations (in LF coordinate $x^-$).
The vacuum state can be identified with the eigenstate of the $P_-$ with minimal
eigenvalue $p_-=0$.

The field operator $\Phi(x)$ at $x^+=0$ can be classified in $p_-$
via the following form of Fourier
decomposition (here the $\Phi(x)$ is taken to be a scalar field):
 \disn{1}{
\Phi(x)=\int\limits_0^\infty\frac{dp_-}{\sqrt{2p_-}}
\ls a(p_-,x^\p)e^{-ip_-x^-}+{\rm h.c.}\rs,
\nom}
where $a(p_-,x^\p)$ and $ia^+(p_-,x^\p)$
at $p_->0$ enter into scalar field action as
canonically conjugated variables on the LF:
 \disn{2}{
[a(p_-,x^\p),a({p'}_-,{x'}^\p)]=0,
\nom}
 \disn{3}{
[a(p_-,x^\p),a^+({p'}_-,{x'}^\p)]=\de(p_--{p'}_-)\de(x^\p-{x'}^\p).
\nom}
The LF momentum operator is
 \disn{4}{
P_-=\int dx^-\int d^2x^\p(\dd_-\Phi)^2=
\int d^2x^\p\int\limits_0^\infty
dp_- p_- a^+(p_-,x^\p)a(p_-,x^\p)\ge 0.
\nom}
"Physical vacuum"  $|0\rangle$ is defined like "mathematical" one:
 \disn{5}{
a(p_-,x^\p)|0\rangle=0,\quad p_->0.
\nom}
This simplicity of the description of the vacuum is main advantage of
the quantization on the LF.

{\bf 2.}   The problem of searching the spectrum of bound states
can be considered
nonperturbatively in LF Fock space basis $\{a^+....a^+|0\rangle\}$
by solving the following equations:
 \disn{6}{
P_+|\Psi\rangle =p_+|\Psi\rangle,\quad
P_-|\Psi\rangle=p_-|\Psi\rangle,\quad
P_{\p}|\Psi\rangle=0.
\nom}
Then $m^2=p^2=2p_+p_-$.
This can be used in attempts to approach nonperturbatively
to bound state problem in Quantum Chromodynamics (QCD).

{\bf 3.}  Main difficulties of LF quantization are related
with singularities at $p_-\to 0$.

Possible regularizations are

{\bf (a)} cutoff in $p_-\;\; (p_-\ge\e>0)$, or

{\bf (b)} the "DLCQ" regularization ("DLCQ" means Discretized
Line Cone Quantization), i.e. the cutoff in $x^-$
($|x^-|\le L$ plus  periodic boundary conditions, and $p_-=\frac{\pi n}{L}$,
$n=0,1,2,\dots$). The $p_-=0$ mode in the DLCQ is to
be expressed canonically through other modes
(for gauge theory this was studied by Novozhilov, Franke,
Prokhvatilov \cite{nov2,nov2a}).

Both types of regularization can break Lorentz symmetry.
This destroys usual
perturbative renormalization.
The problem of restoring the symmetries, broken by the regularization on the LF,
and proving the equivalence of usual and LF quantizations is rather difficult.
Nevertheless it can be solved, at least
perturbatively \cite{burlang} (and to all orders in coupling
constant \cite{tmf97}) via comparison of two Feynman perturbation theories: one
generated by usual and one by LF quantizations.
Such a comparison shows the necessity of adding
to regularized canonical LF Hamiltonian unusual "counterterms"
which restore the mentioned equivalence in perturbation theory
in the limit of removing the regularization.

The "zero" modes (i.e. $p_-=0$ modes) and
modes with $p_-$ in the vicinity of $p_-=0$
may be important for the
description of nonperturbative vacuum effects like condensates.
These vacuum effects can be introduced semiphenomenologically
using as a guide solutions of simplified models.
As an example we will consider implications of QED(1+1) (massive Schwinger
model)  for such a simplified description of vacuum effects
in more complicated cases.
Attempts to use this model as a guide are
common \cite{jackiw}.

This model has gauge symmetry, nontrivial topological effects and confinement
 of fermions like QCD. Furthermore, it has "dual" description in terms of scalar
boson field, and this description allows to see vacuum effects, nonperturbative
from the point of view of usual QED(1+1) coupling.

\section{QED(1+1) on the Light Front}
The Lagrangian density is
 \disn{7}{
L=-\frac{1}{4}F_{\m\n}F^{\m\n}+\bar\psi(i\g^\m D_\m-M)\psi,
\nom}
where $\m,\n=0,1$,
$F_{\mu\nu}= \dd_{\mu} A_{\nu}- \dd_{\nu }A_{\mu}$,
$D_{\mu}=\dd_{\mu} - i\eel A_{\mu}$,
 \disn{8}{
\g^0=\ls
\begin{array}{cc}
0 & -i\\
i & 0
\end{array}
\rs\! , \quad \g^1=\ls
\begin{array}{cc}
0 & i\\
i & 0
\end{array}
\rs\! .
\nom}

Canonical Hamiltonian on the LF in $A_-=0$ gauge is
 \disn{9}{
P_+=\int dx^-\biggl(\frac{\eel^2}{2}\ls \dd_-^{-1}
[{\psi_+}^+\psi_+]\rs^2-
\frac{iM^2}{2}{\psi_+}^+\dd_-^{-1}\psi_+\biggr),
\nom}
where the $\psi=\ls {\psi_+ \atop \psi_-} \rs$
and the $F_{+-}= -\dd_-A_+ $ are already expressed owing to
canonical constraints, as follows:
 \disn{10}{
F_{+-}=-\sqrt{2}\,e\,\dd_-^{-1}\ls {\psi_+}^+\psi_+\rs,\quad
\psi_-=\frac{M}{\sqrt{2}}\,\dd_-^{-1}\psi_+.
\nom}
Feynman perturbation theory in coupling constant $e$ has
strong infrared divergences. So we began our study from boson form.
The transition to this boson form can be performed in different,
but equivalent ways.

The result can be described by the Lagrangian density
 \disn{11}{
L=\frac{1}{2}\dd_\m\f\dd^\m\f-\frac{1}{2}m^2\f^2+
\frac{Mme^C}{2\pi}:\cos\ls \te+\sqrt{4\pi}\f\rs:,
\nom}
where $m=\eel/\sqrt\pi$,
the boson field $\f(x)$ can be related to fermion current,
the $\theta$ is so called "$\theta$"-vacuum parameter,
$C$ is the Euler constant, and $:\ \ :$ means normal ordering
in interaction picture.
Fermion mass $M$ (in fact, dimensionless parameter $M/e$) plays the role of
coupling constant.

We quantized this boson theory on the LF
with the $p_-\ge\e>0$ regularization and considered the
difference between LF perturbation theory and corresponding
covariant one (in Lorentz coordinates) to all
orders in the $M$ \cite{tmf02}.
The found difference can be generated by the "counterterms",
which  must be added to canonical LF Hamiltonian.
The LF Hamiltonian corrected in this way has the form:
 \disn{13}{
P_+=\!\int\! dx^-\!\ls\frac{1}{2}\,m^2:\f^2:
-\frac{Mme^C}{2\pi}:\cos\ls \hat\te+\sqrt{4\pi}\f\rs:\rs-\ns-
\!\int\! dx^-\!\int\! dy^-\,
\frac{M^2}{2\pi\,|x^--y^-|}\ls
:e^{i\sqrt{4\pi}\f(x^-)}e^{-i\sqrt{4\pi}\f(y^-)}:-1\rs.
\nom}
where the $\hat\te$ is the parameter, replacing initial $\te$.
It is related with fermion condensate
and will be  specified below.

One can return again to fermion field
variables on the LF, using DLCQ type of the regularization with
$|x^-|\le L$ and antiperiodic in $x^-$
boundary condition for fermion field.
As in \cite{tmf02} we identify
fermion field $\psi_+(x)$ describing unconstrainted component $\psi_+(x)$
of fermion field $\Psi=\ls {\psi_+\atop \psi_-}\rs$
on the LF with following expression:
 \disn{14}{
\psi_+(x)=\frac{1}{\sqrt{2L}}\;e^{-i\om}e^{-i\frac{\pi}{L}x^-(Q-\frac{1}{2})}
:e^{-i\sqrt{4\pi}\f(x)}:.
\nom}
Remark again that the $\psi_+(x)$ is chosen to be
antiperiodic while $\f(x)$ is taken periodic in $x^-$
without zero mode.

The $Q$ is the "charge" operator on the LF:
 \disn{15}{
Q=\sqrt{2}\int\limits_{-L}^Ldx^-\,:{\psi_+}^+(x)\psi_+(x):.
\nom}
The $\om$ is the variable canonically conjugated to $Q$:
 \disn{16}{
[\om,Q]_{x^+=0}=i,\qquad
e^{i\om}Qe^{-i\om} =Q+1.
\nom}
We can define this operator more exactly.
We introduce Fourier decomposition for the $\psi_+(x)$:
 \disn{18}{
\psi_+(x)=\frac{1}{\sqrt{2L}}\ls \sum_{n\ge 1}b_n
e^{-i\frac{\pi}{L}(n-\frac{1}{2})x^-}+ \sum_{n\ge 0}d_n^+
e^{i\frac{\pi}{L}(n+\frac{1}{2})x^-}\rs\! ,
\nom}
where at $x^+=0$
 \disn{19}{
\{b_n,b_{n'}^+\}=\{d_n,d_{n'}^+\}=\de_{nn'},
\{b_n,b_{n'}\}=\{d_n,d_{n'}\}=0,
\nom}
because the expression (\ref{14}) satisfies canonical
anticommutation relations for fermion fields on the LF.

We define the vacuum $|0\ra$ as a state corresponding
to the minimum of the $P_-$:
 \disn{20}{
P_- =\sum_{n\ge 1}b_n^+b_n\frac{\pi}{L}\ls n-\frac{1}{2}\rs+
\sum_{n\ge 0}d_n^+d_n\frac{\pi}{L}\ls n+\frac{1}{2}\rs.
\nom}
Hence,
 \disn{21}{
b_n|0\ra=d_n|0\ra=0.
\nom}
For the charge $Q$ we get
 \disn{22}{
Q=\sum_{n\ge 1}b_n^+b_n-\sum_{n\ge 0}d_n^+d_n.
\nom}
We can  fix the operator $e^{i\om}$ as follows:
 \disn{23}{
e^{i\om}\psi(x)e^{-i\om} =e^{i\frac{\pi}{L}x^-}\psi(x),
\qquad \psi_n\to\psi_{n+1},
\nom}
 \disn{24}{
e^{i\om}|0\ra=b^+_1 |0\ra,\quad e^{-i\om}|0\ra=d^+_0 |0\ra,
\nom}
that agrees with the definition of vacuum state $|0\ra$
as filled Dirac sea:
 \disn{24.1}{
|0\ra=d_0d_1\dots|0_D\ra,\qquad
\Psi(x)|0_D\ra=0.
\nom}
One can show \cite{tmf02} that the boson form of the corrected
LF Hamiltonian $P_+$ transforms to following fermion form:
 \disn{25}{
P_+=\int\limits_{-L}^Ldx^-\biggl(\frac{\eel^2}{2}\ls \dd_-^{-1}
[{\psi_+}^+\psi_+]\rs^2-
\frac{iM^2}{2}{\psi_+}^+\dd_-^{-1}\psi_+
-\ns
-\frac{eMe^C\sqrt{2L}}{4\pi^{3/2}}\ls
e^{-i\hat\te(M/\eel,\:\te)-i\frac{\pi}{2L}x^-}
\:e^{i\om}\psi_+ +h.c.\rs\biggr)=\no
=\int\limits_{-L}^Ldx^-\biggl(\frac{\eel^2}{2}\ls \dd_-^{-1}
[{\psi_+}^+\psi_+]\rs^2-
\frac{iM^2}{2}{\psi_+}^+\dd_-^{-1}\psi_+
-\frac{eMe^C}{4\pi^{3/2}}\ls
e^{-i\hat\te(M/\eel,\:\te)}
\:e^{i\om}d_0^+ +h.c.\rs\biggr).
\nom}

The $\hat\theta(M/e,\te)$ is related to fermion
condensate parameters \cite{tmf02}:
 \disn{26}{
\sin\hat\te =\frac{2\pi^{3/2}}{e\:e^C}
\langle\Omega|:\bar\psi\g^5\psi:|\Omega\ra=
\langle\Omega|:\sin(\f+\te):|\Omega\ra,
\nom}
where the $|\Omega\ra$, $\psi$ and $\f$ mean physical vacuum,
fermion field  and corresponding boson field in the theory,
quantized in Lorentz coordinates.

It is possible to show that
 \disn{27}{
\hat\theta(M/e,-\te)=-\hat\theta(M/e,\te),\qquad
\hat\theta(M/e,0)=0,\qquad
\hat\theta(M/e,\pi)=\pi.
\nom}

It is remarkable that the "counterterm"
(which restores the equivalence with formulation in Lorentz
coordinates
in boson form of $P_+$), being proportional to $M^2$, exactly
coincides with corresponding (proportional to $M^2$)
 canonical term of the $P_+$ in the
fermion form. The terms, proportional to $Me^{\pm i\hat\te}$
are not present in canonical fermion form of LF Hamiltonian. They depend
on vacuum condensate parameter $\hat\te$ through fermion
"zero" modes $d_0,d_0^+$ which enter the Hamiltonian linearly
via "neutral" combinations $e^{i\om}d_0^+$ and $d_0e^{-i\om}$.

Any bound state with $Q=0$ and fixed finite value of the
$p_-=(\pi K)/L$ can be described in terms of fermion Fock space
(formed with $b^+d^+$ acting on the vacuum $|0\ra$). Owing to the
positivity of the spectrum of the $P_-$ for these states (which are taken
to be orthogonal to the vacuum),
the Hamiltonian $P_+$ on this subspace can be represented
by finite dimensional matrix.

One can calculate the spectrum of mass
 \disn{29}{
m^2=2p_-p_+=\frac{2\pi K}{L}p_+
\nom}
numerically for different integer $K$.
An extrapolation of results to $K\to\infty$ gives "true" values of
mass.

\begin{figure}[hb]
\begin{center}
\epsfig{file=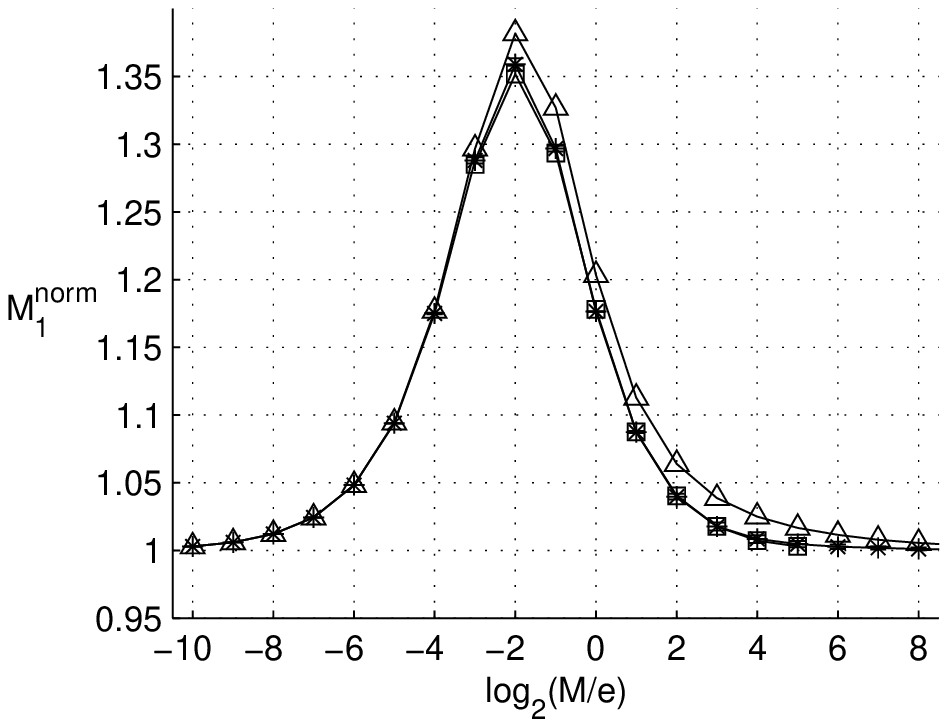,height=10cm}
\caption{
The results of the calculation
of the mass $M_1$ of lowest bound state at $\hat\te=\te=0$;
$*$~corresponds to results, obtained by the extrapolation to the domain
$K\to\infty$,
$\bigtriangleup$~corresponds to $N=30$,
$\square$~corresponds to known results of the calculation \cite{ham1}
on the lattice.
}
\end{center}
\end{figure}

We have found good agreement \cite{yf05} with lattice calculations
in Lorentz coordinates
for $\theta=0$ at any $M/e$, see fig.~1, and for $\theta=\pi$
at small $M/e$ (for larger $M/e$ we see the behaviour of the
spectrum, indicating possible phase transition, that is seen also
in mentioned lattice calculations), see fig.~2.
On the fig.~1 we use normalized value
 \disn{30}{
M^{\rm norm}_1=\frac{M_1}{\sqrt{m^2+(2M)^2}},
\nom}
where $M_1$ is the mass of lowest bound state.

\begin{figure}[t]
\begin{center}
\epsfig{file=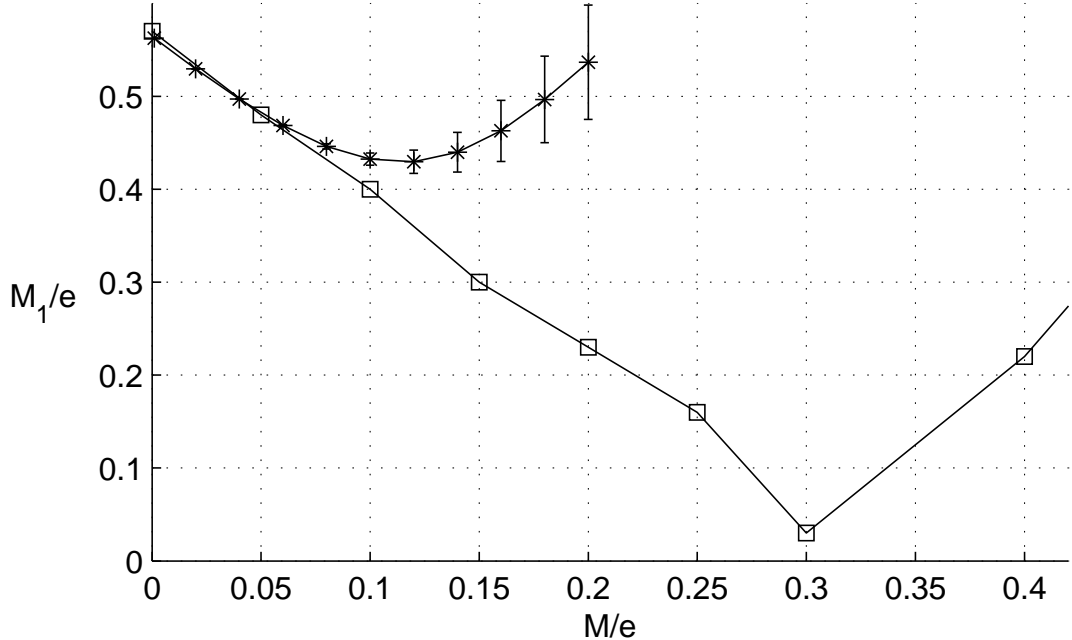,height=9cm}
\caption{
The results of the calculation
of the mass $M_1$ of lowest bound state at $\hat\te=\te=\pi$;
$*$~corresponds to results, obtained by the extrapolation to the domain
$K\to\infty$,
$\square$~corresponds to known results of the calculation \cite{ham2}
on the lattice.
}
\end{center}
\end{figure}

Beside of that we have calculated the spectrum for any other
values of the parameter $\hat\theta$ (or for $0<\te<\pi$)
that was not yet done with
lattice in Lorentz coordinates. We have found for all nonzero
$\hat\theta$ some "critical" region  of $M/e$ where the  mass
spectrum becomes unbounded from the bottom. This indicates
that the perturbation theory in $M/e$, that we used
in our analysis, fails at these "critical" values of $M/e$.

It is very interesting problem to find a way to continue our
LF Hamiltonian to "nonperturbative" in $M/e$ region.

Let us consider only  LF Hamiltonian obtained  perturbatively to
all orders in $M/e$.
We can formally  find some Lagrangian, that generates  this Hamiltonian
via
canonical formulation on the LF. Assuming that this Lagrangian is
the ordinary one plus some counterterms, it is easy to find that
these counterterms must be such that         only
the constraint, which connects $\psi_-$ and $\psi_+$
 components of
bispinor field $\psi=\ls {\psi_+ \atop \psi_-} \rs$, should be
modified.
In naive canonical formulation of QED(1+1 ) on the LF such constraint
has the following form:
 \disn{31}{
\sqrt{2}\,\dd_-\psi_--M\psi_+=0
\nom}
 being the one of components of Dirac equation.  "Zero" mode of the
$\psi_-$ is unconstrainted by this equation.

For our DLCQ formulation with
antiperiodic in $x^-$ fields $\psi$ we introduce the following modification
of this constraint :
 \disn{32}{
\sqrt{2}\,\dd_-\psi_--M\psi_++\frac{e\:e^C}{2\pi^{3/2}}
e^{i(\omega-\hat\te)+i\frac{\pi}{2L}x^-}=0.
\nom}
Here the $\psi_-$ has no zero mode because of antiperiodic
boundary conditions.

Now one can construct modified  QED(1+1) Lagrangian,
that generates this form of the constraint. Having such a Lagrangian
one can construct corresponding LF Hamiltonian and substitute the solution
of the constraint w.r.t. the $\psi_-$ into  this LF Hamiltonian.
In this way we come back to abovementioned form of  corrected
LF Hamiltonian.

The expression for the $\psi_-$, respecting our modified
 constraint, includes
the operator $e^{i\omega}$. Taking into account the properties
 of this operator, we can check that  correct values of
 vacuum condensate parameters, can be get on the LF
using  LF vacuum $|0\ra$ for  corresponding VEVs:
 \disn{33}{
\left.\langle 0|{\psi_-}^+\psi_+|0\ra\right|_{M\to0}=
\left.\langle 0|\frac{1}{\sqrt{2}}\,\dd_-^{-1}\ls
M{\psi_+}^+-\frac{e\,e^C}{2\pi^{3/2}}e^{-i(\om-\hat\te)-i\frac{\pi}{2L}x^-}\rs
\psi_+|0\ra\right|_{M\to0}\!\!\!
=\frac{ee^Ce^{i\hat\te}}{2\pi^{3/2}}.
\nom}

\section{ A description of vacuum condensate in QCD on the LF}
The remark at the end of previous section implies a
possible way of semiphenomenological
introduction of vacuum parameters in QCD(3+1) on the LF.

Namely, we can assume similar modification of the
(3+1)-dimensional analog of the constraint equation, relating the
$\psi_-$ and the $\psi_+$ on the LF,
using the same DLCQ
formulation.
To make this correctly one has to introduce a lattice
with respect to transverse coordinates. Here we only
describe the idea. We introduce (3+1)-dimensional analog
of operators $e^{i\om}$ and $Q$, defining for each component of quark
field $\psi_j(x)$ some unitary operator $U_j(x^\p)$  and
"transverse charges" $Q_j(x^\p)$.
We require
 \disn{34}{
U_j(x^\p)\psi_j(x){U_j}^+(x^\p) =e^{i\frac{\pi}{L}x^-}\psi_j(x).
\nom}
Beside of that
 \disn{35}{
[U_j(x^\p),Q_{j'}({x'}^\p)]=\de_{jj'}\de_{x^\p{x'}^\p}U_j(x^\p),\cr
Q_j(x^\p)|0\ra=0,\cr
U_j(x^\p)|0\ra=d_{0j}^+(x^\p)|0\ra,\cr
U_j^+(x^\p)|0\ra=b_{1j}^+(x^\p)|0\ra.
\nom}

Then we can write the component of Dirac equation that presents
our LF constraint as follows:
 \disn{36}{
\sqrt{2}\,\dd_-\psi_{-j}(x^\p)+(\hat D_\p-M)\psi_{+j}+
\varkappa U_j(x^\p)e^{i\frac{\pi}{2L}x^-}=0,
\nom}
where
 \disn{37}{
\hat D_\p=\sum_{k=1,2}\sigma_kD_k,
\nom}
and the $\varkappa$ is a parameter related with possible vacuum
effects.
Resolving this constraint w.r.t. $\psi_{-j}$ we can obtain
modified form of the QCD LF Hamiltonian, including
vacuum parameter $\varkappa$, and consider
bound state problem with this Hamiltonian.

First results on this way were discussed by S. Dalley and
G. McCartor \cite{mccardal}.
They showed that new terms influence the $q\bar q$ meson mass
spectrum, correctly splitting masses of $\pi$ and $\rho$ mesons.

\vskip 1em
{\bf Acknowledgments.}
This work was supported by Russian Federal Education
Agency, Grant No.~RNP.2.1.1.1112,
and President of Russian Federation, Grant No.~NS-5538.2006.2.
The work was supported also in part (S.~A.~P. and E.~V.~P.) by the Russian
Foundation for Basic Research, Grant No.~05-02-17477.

\end{document}